\documentclass[aps,prb,reprint,superscriptaddress]{revtex4-1}

\usepackage[colorlinks=true,citecolor=blue,menucolor=blue,linkcolor=blue,urlcolor=blue]{hyperref}

\usepackage{amsfonts}
\usepackage{amssymb}
\usepackage{amsmath}
\usepackage{longtable}
\usepackage{graphicx}
\usepackage[tabke]{xcolor}
\usepackage{multirow}
\usepackage{color}
\usepackage{physics}

\usepackage[normalem]{ulem}

\newcommand{\angstrom}{\text{\normalfont\AA}}

\begin{document}

\title{Plane-wave many-body corrections to the conductance in bulk tunnel junctions}

\author{Alberto Dragoni}
\email{alberto.dragoni@cea.fr}
\affiliation{Univ. Grenoble Alpes, 38000 Grenoble, France}
\affiliation{CEA, LETI, MINATEC Campus, 38054 Grenoble, France}
\affiliation{CNRS, Institut N\'{e}el, 38042 Grenoble, France}

\author{Beno\^{i}t Skl\'{e}nard}
\email{benoit.sklenard@cea.fr}
\affiliation{Univ. Grenoble Alpes, 38000 Grenoble, France}
\affiliation{CEA, LETI, MINATEC Campus, 38054 Grenoble, France}

\author{Valerio Olevano}
\email{valerio.olevano@neel.cnrs.fr}
\affiliation{Univ. Grenoble Alpes, 38000 Grenoble, France}
\affiliation{CNRS, Institut N\'{e}el, 38042 Grenoble, France}

\author{Fran\c{c}ois Triozon}
\email{francois.triozon@cea.fr}
\affiliation{Univ. Grenoble Alpes, 38000 Grenoble, France}
\affiliation{CEA, LETI, MINATEC Campus, 38054 Grenoble, France}

\date{\today}

\begin{abstract}
The conductance of bulk metal--insulator--metal junctions is evaluated by the Landauer formula using an \textit{ab initio} electronic structure calculated using a plane-waves basis set within density-functional theory (DFT) and beyond, i.e.\ including exact non-local exchange using hybrid functional (HSE) or many-body $G_0W_0$ and COHSEX quasiparticle (QP) schemes.
We consider an Ag/MgO/Ag heterostructure model and we focus on the evolution of the zero-bias conductance as a function of the MgO film thickness. Our study shows that the correction of the electronic structure beyond semi-local density functionals goes in the right direction to improve the agreement with experiments, significantly reducing the zero-bias conductance. This effect becomes more evident at larger MgO thickness, that is in increasing tunneling regime. We also observe that the reduction of the conductance seems more related to the correction of wavefunctions rather than energies, and thus not directly related to the correction of the band-gap of bulk MgO. $G_0W_0$ and HSE both provide a correct band-gap in agreement with experiments, but only HSE gives a significant reduction of the conductance. COHSEX, while overestimating the band-gap, gives a reduction of the conductance very close to HSE.
\end{abstract}

\maketitle

\section{Introduction}
Metal--insulator--metal junctions are of high technological interest since they are used in currently developed electronic devices, like new generation memories based on the variation of the electric resistance (RRAM\cite{OxRAM_TS,cbram_abinitio}), or magnetoresistive devices \cite{Butler_abinitio_Fe-MgO_2001,mathon_magnetoresistance, Gangineni_Fe-MgO_2014}.
First-principles calculations of the tunneling conductance in these junctions can provide an important understanding to drive technological developments.
With the downscaling of microelectronics, the typical size of devices can reach a few nanometers, which calls for an atomistic and quantum treatment of the electronic transport. In particular, the calculation of the transmission of electrons through an interface between two materials requires an accurate description of the electron wavefunctions at the atomic scale.

The prevalent approach to study quantum transport at the atomic scale is the \textit{ab initio} ``DFT+Landauer'' scheme. Density Functional Theory (DFT)\cite{HohenbergKohn64,KohnSham65} is used to calculate the electronic structure of the junction and of the leads. The probability of transmission of electrons through the junction is then calculated using the Green's function method, and the zero-bias conductance is obtained via the multi-channel Landauer formula\cite{Landauer57,Landauer70,Buttiker85,Datta}. Several available codes, e.g.\ the ones documented in Refs.~[\onlinecite{PapiorTS,WannierTransport,Mostofi_CPC_2008}], implement DFT+Landauer on spatially localized basis sets (atomic orbitals, Wannier functions), which are more convenient to compute the quantum transmission.
 
This approach has two main limitations. Firstly the Landauer formalism only holds in the coherent regime of transport, which means that inelastic scattering such as electron-electron and electron-phonon interactions cannot be taken into account. Secondly only calculations close to the zero-bias limit are formally correct, since so far there exist no exact formulation of DFT for out-of-equilibrium and open systems (i.e. varying number of electrons).
However some DFT+Landauer codes, such as \textsc{TranSiesta}\cite{PapiorTS}, allow applying a finite bias voltage between the electrodes, still using the equilibrium density functional. Despite the incompleteness of such approach\cite{landauer_incompleteness}, it allows addressing realistic devices within a reasonable computing time and provides valuable information about the transmission of electrons through interfaces, for instance when the symmetries of the wavefunctions in each material strongly impact the transmission\cite{Butler_abinitio_Fe-MgO_2001}. The DFT+Landauer method should also provide a good estimation of the elastic scattering of electrons by defects and impurities. This holds in the coherent regime, which is a reasonable approximation for short junctions (a few nanometers) and at very low temperature, so that the electrons do not loose their phase coherence between the two electrodes.

Unfortuntately, even in the coherent regime and at zero bias voltage, DFT+Landauer turns out to catch only a qualitative description of transport, and to systematically overestimate conductances by one or even several order of magnitudes in the tunneling regime. This was pointed out by Vignale and Di Ventra\cite{landauer_incompleteness}, and there are several hypothesis about this failure. One of them\cite{Neaton,Quek,Mowbray-Thygesen} attributes the systematical overestimation of the conductance to the systematical underestimation of the HOMO-LUMO gap in molecules, and analogously of the band-gap in insulators, by DFT\cite{Perdew_bandgap}.

This works aims at overcoming this particular DFT+Landauer limitation, in the coherent regime and at zero bias, by going beyond DFT towards many-body quasiparticle (MB-QP) approaches, including the $GW$\cite{HedinGW} or the COHSEX\cite{HedinGW} (Coulomb-hole screened exchange) approximations.
We also consider the Heyd-Scuseria-Ernzerhof (HSE) hybrid functional\cite{BeckeHSE93} which, rigorously speaking, cannot be considered an approximation within DFT like the LDA or GGA, and should be rather considered as a semi-empirical MB approximation, with a modified Hartree-Fock exchange Fock operator mimicking the COHSEX statically screened exchange non-local self-energy. With respect to DFT LDA or GGA, these QP approaches certainly provide a more physical electronic structure, and this should translate into an improved description of quantum transport properties. The approach used in this work is still based on the Landauer formalism, but with a corrected electronic structure. Along these lines several authors have already attempted to go beyond DFT+Landauer to describe quantum transport in molecular junctions, for example by using semi-empirical models like the image-charge DFT+$\Sigma$ model\cite{Quek}, or fully \textit{ab initio} approaches\cite{Olevano2011,Olevano2017} like $GW$ or COHSEX. Such attempts remain rare in the literature. The main reason is that MB-QP calculations on bulk materials are not implemented in the available DFT codes using localized basis sets, which are the most suitable for transport calculations. Moreover, MB-QP corrections on bulk tunnel junctions, such as the Ag/MgO/Ag heterostructure studied here, are unexplored so far. By bulk junctions, we mean junctions with macroscopic dimensions along the directions parallel to the interfaces.

By relying on a quantum transport implementation based on plane-waves and Wannier functions (see section \ref{sec:method}), we compute the quantum transmission spectra and zero-bias conductances of metal--insulator--metal junctions made of a thin film of MgO sandwiched between two bulk Ag electrodes. This model of heterostructure is reasonable since, experimentally, thin layers of MgO can be epitaxially deposited on an Ag substrate\cite{Schintke2001}. Considering ferromagnetic electrodes would be of higher technological interest as they form magnetic tunnel junctions with MgO\cite{Butler_abinitio_Fe-MgO_2001,Gangineni_Fe-MgO_2014,mathon_magnetoresistance}. However, we have chosen Ag electrodes to avoid the additional complexity of spin polarized transport and to focus only on the effect of MB corrections on the tunneling conductance.

We study the transport properties of the junction as a function of the oxide thickness, from 1 to 4 MgO cubic cells, and we consider four different approximations: DFT in the GGA PBE approximation; $G_0W_0$\cite{HedinGW} on top of DFT, which takes into account only a diagonal $GW$ self-energy and allows to recalculate only the QP energies, keeping the DFT wavefunctions unchanged; the self-consistent COHSEX approximation, which instead recalculates both QP energies and wavefunctions; the HSE hybrid functional, which also updates both energies and wavefunctions. In section~\ref{sec:results}, we present the simulation results and we analyze the effect of the different approximations on the transport properties in the tunneling regime.

\section{Method and computational details} \label{sec:method}
In this section we describe the method and the computational details used to calculate the tunneling conductance through Ag/MgO/Ag junctions. The procedure is divided into three parts.
\begin{enumerate}
\item The electronic structure of Ag/MgO/Ag supercells is computed self-consistently on a plane-waves basis set. The calculations are performed within DFT or using various MB corrections on top of DFT in order to compare their impact on electronic transport.
\item A basis set transformation is performed from plane-waves to maximally localized Wannier functions\cite{Marzari1997,Marzari2001,Marzari2012}, using the \textsc{Wannier90} code\cite{Mostofi_CPC_2014}. The Hamiltonian expressed in this basis is convenient for transport calculations.
\item The electron transport through Ag/MgO/Ag junctions is computed by the Landauer formalism\cite{Buttiker85} and the Green's functions method, using our in-house code which allows calculating the transmission for any transverse wavevector $\mathbf{k}^\parallel$ parallel to the Ag/MgO interfaces. This is necessary for evaluating the conductance through junctions with macroscopic surface area.
\end{enumerate}
\subsection{DFT computational details}\label{subsec:dft}
\textit{Ab initio} calculations based on DFT are carried out using the \textsc{VASP} code\cite{Kresse_PRB_1996,Kresse_PRB_1999} with the PBE\cite{PBE96} functional. The core-valence interaction is described by Projector Augmented Waves (PAW) datasets including 4\textit{d} and 5\textit{s} states for Ag, 2\textit{s} and 2\textit{p} for O, and 3\textit{s} for Mg. Electron wavefunctions are expanded in a plane-waves basis set with a kinetic energy cut-off of 415~eV. HSE calculations and MB corrections are detailed in section \ref{subsec:hse_mb}.\\
For comparison purpose, DFT calculations have also been performed using the \textsc{Siesta} code\cite{Ordejon96,Soler02}. The chosen functional is also PBE, while the pseudopotentials are of norm-conserving Troullier-Martins type\cite{Troullier1993} and include the same states as the \textsc{VASP} PAW datasets. A polarized double-zeta basis set of atomic orbitals is used, with a cut-off radius yielding an energy shift of 25 meV. The plane-waves cut-off for the real space grid is taken equal to 450 Ry.

\subsection{Structures studied in \textit{ab initio}}
Ag and MgO both have face-centered cubic structure and are stacked along the [100] direction. O atoms are located in front of Ag atoms, which is known to be the most stable configuration \cite{Schintke2001}. In the following, $m$Ag/$n$MgO/$m$Ag denotes a supercell with $n$ MgO cubic cells surrounded on both sides by $m$ Ag cubic cells.

Cell volume relaxations are performed on bulk Ag and bulk MgO, yielding lattice constants of $4.14$ \angstrom \ and $4.20$ \angstrom, respectively. Then structural relaxations of 2Ag/$n$MgO/2Ag supercells are performed, $n$ ranging from 1 to 4. The Brillouin zone is sampled using a $1 \times 4 \times 4$ Monkhorst-Pack (MP) grid. The lattice constants along directions [010] and [001], parallel to the interfaces, are fixed to the lattice constant of bulk Ag ($4.14$ \angstrom) in order to mimick thin layers of MgO epitaxially sandwiched between thick unstrained Ag layers. Hence the MgO layer is compressively strained in the in-plane directions. The length of the supercell along [100] is relaxed.
The interfacial distance turns out to be $d \simeq 2.69$ \angstrom \  between the Ag and MgO atomic planes, in good agreement with previous studies\cite{Schintke2001}. The MgO lattice constant is elongated to about $4.33$ \angstrom \ along [100].

\begin{figure}[t]
\centering
\includegraphics[width=\linewidth]{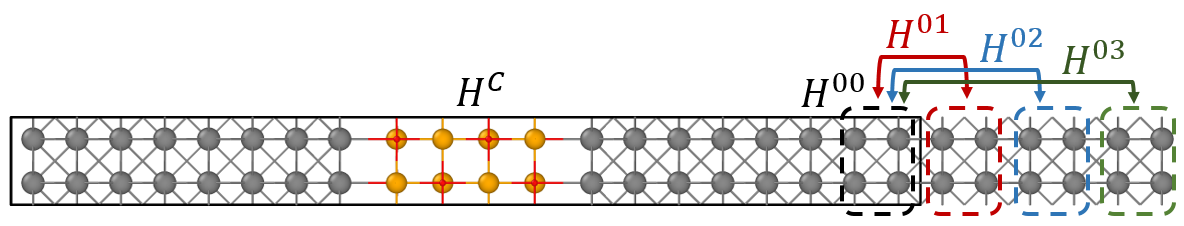}
\caption{4Ag/2MgO/4Ag supercell (outlined by a solid line rectangle) studied in \textit{ab initio}. 3 Ag cubic cells of its right-side periodic replica are also represented. The Hamiltonian blocks used for transport calculations are schematized: $H^C$ is the Hamiltonian of the central region and the blocks $H^{0i}$ define the semi-infinite periodic Ag electrodes. See section \ref{subsec:conductance} for details.}
\label{structure}
\end{figure}

Supercells with thicker Ag layers are needed to define the semi-infinite periodic electrodes in transport simulations (see Fig.~\ref{structure} and section \ref{subsec:conductance}). They are built from the relaxed systems discussed above, by keeping the relaxed atomic positions of the MgO layer and of 1 cubic cell of Ag on each side of MgO, and by adding 3 Ag cubic cells on each side, with an inter-plane distance fixed to the bulk value, $2.07$ \angstrom. This ensures the perfect periodicity of the first 3 and last 3 Ag unit cells, which will be used to define the Hamiltonian of the semi-infinite electrodes. In the following, all \textit{ab initio} calculations (DFT, HSE, and MB) are performed on these 4Ag/$n$MgO/4Ag supercells, with a $2 \times 4 \times 4$ MP grid containing $\Gamma$.

\subsection{HSE calculations and many-body corrections}
\label{subsec:hse_mb}

HSE calculations are carried out using \textsc{VASP} with the HSE06 hybrid functional\cite{HSE06}. According to Vi\=nes \textit{et al.}\cite{HSEstudyMgO}, HSE06 requires an $\alpha$ mixing parameter higher than 0.4 to reproduce the MgO experimental gap. Here $\alpha$ is fixed to 0.43, which gives an electronic band-gap at $\Gamma$ of 7.77~eV for the MgO primitive cell, in agreement with the experimental value which lies between 7.67 and 7.83~eV\cite{HSEstudyMgO}. The screening parameter is fixed to 0.2 \angstrom$^{-1}$.

MB effects are accounted for by using $G_0W_0$ (correction of eigenvalues only) and self-consistent COHSEX (correction of both eigenvalues and eigenfunctions) on top of DFT. The use of a full self-consistent $GW$ calculation on both eigenvalues and eigenfunctions is not considered here, due to the high computational cost. COHSEX is an approach half-way between Hartree-Fock\cite{SlaterHF} and $GW$. The HF bare exchange is statically screened in COHSEX, and it considers an additional ``Coulomb hole'' term which accounts for a classical response of the medium after the addition of a point charge (electron or hole). It requires less computational resources than $GW$, but it systematically overestimates the band-gap. For $G_0W_0$ calculations, we use a cut-off energy of 210~eV for the response function, and 64 frequency grid points. The number of empty bands used in the dielectric function and for the Green's function is 6144 for the structures with 1 and 2 MgO cubic cells, and 6656 for the structures with 3 and 4 MgO cubic cells. For self-consistent COHSEX the number of bands is reduced to 2048, except for the biggest structure (4 MgO), which uses 3072 bands. The number of self-consistent cycles is taken equal to 3.

\subsection{Transformation to a Wannier basis set}
\label{subsec:wannier}

The transformation from Bloch waves to maximally localized Wannier functions is performed with the \textsc{Wannier90} code\cite{Mostofi_CPC_2014} and its interface with \textsc{VASP}. The bands around the Fermi level are not isolated in the energy spectrum. Hence we use the method of Souza \textit{et al.}\cite{Marzari2001}: the bands within a "frozen" energy window are completed by an optimal subspace of higher energy bands. The Wannier functions describe exactly the bands within the frozen window, which is chosen here between the bottom of the valence band up to 3 eV above the Fermi level.

A set of trial orbitals is needed as a starting point for the \textsc{Wannier90} code. For Ag, we choose the set of orbitals proposed by Souza \textit{et al.}\cite{Marzari2001} for copper: 5\textit{d} orbitals centered on each atom, and interstitial \textit{s} orbitals. For MgO, we choose 1\textit{s} and 3\textit{p} orbitals on oxygen atoms, and also interstitial \textit{s} orbitals. Hence there are 28 Wannier functions per Ag cubic cell and 24 Wannier functions per MgO cubic cell, which yields 320 Wannier functions for the largest structure (4Ag/4MgO/4Ag). The number $N_b$ of \textit{ab initio} bands included in the optimization of the Wannier subspace is chosen at least twice the number $N_w$ of Wannier functions. Using 100 iterations for optimizing the subspace ("disentanglement" step) and 100 iterations for optimizing the Wannier functions, we obtained well-localized functions with quadratic spreads not exceeding 2 \angstrom$^2$.

We give here some notations and recall some basic properties of Wannier functions \cite{Marzari2012} which will be used in the following. Let $\mathbf{a}_i$ denote the real lattice basis vectors of the \textit{ab initio} supercell, which will be referred to as the "reference supercell" (RS). Here $\mathbf{a}_1$, $\mathbf{a}_2$, and $\mathbf{a}_3$ are chosen along the crystal directions [100], [010], and [001], respectively. Let $\mathbf{g}_i$ denote the reciprocal lattice basis vectors. The \textit{ab initio} Bloch states $\ket{\psi_{n\mathbf{k}}}$ are computed for $\mathbf{k}$-points in a MP grid containing $\Gamma$ with $N_i$ mesh points along each direction $\mathbf{g}_i$. Hence these states satisfy periodic boundary conditions over an extended supercell (ES) made of $N_i$ replicas of the RS along each direction $\mathbf{a}_i$. The replicas are indexed by their translation vector $\mathbf{R}$ with respect to the RS:
\begin{equation}
\mathbf{R} = p_1 \mathbf{a}_1 + p_2 \mathbf{a}_2 + p_3 \mathbf{a}_3
\label{eq:R}
\end{equation}
where $p_i$ are integers. The Wannier functions are optimized linear combinations of Bloch states:
\begin{equation}
\ket{\mathbf{R}n} = \frac{1}{N} \sum_\mathbf{k} e^{-i\mathbf{k} \cdot \mathbf{R}} \sum_{m=1}^{N_b} U_{mn}^{(\mathbf{k})} \ket{\psi_{m\mathbf{k}}}
\label{eq:Wannier_Rn}
\end{equation}
where $n$ indexes the $N_w$ Wannier functions of each cell $\mathbf{R}$, and $N$ is the number of $\mathbf{k}$-points.
The $U^{(\mathbf{k})}$ are $N_b \times N_w$ complex matrices with orthonormal columns, optimized by the \textsc{Wannier90} code to maximize the localization of the Wannier functions.
The average position of $\ket{\mathbf{0}n}$, called Wannier center, is denoted as $\bar{\mathbf{r}}_n$. The phases of the $U^{(\mathbf{k})}$ matrices are chosen so that the $\bar{\mathbf{r}}_n$ are within the RS.
$\ket{\mathbf{R}n}$ is the translation by $\mathbf{R}$ of $\ket{\mathbf{0}n}$, hence its center is at $\bar{\mathbf{r}}_n+\mathbf{R}$.

Wannier functions form an orthonormal basis. They satisfy periodic boundary conditions over the ES. Hence, each Wannier function $\ket{\mathbf{R}n}$ contains periodic replicas centered around positions $\mathbf{R}+\mathbf{T}+\bar{\mathbf{r}}_n$, where $\mathbf{T}$ is any vector of the lattice defined by the periodic repetition of the ES:
\begin{equation}
\mathbf{T} = p_1 N_1 \mathbf{a}_1 + p_2 N_2 \mathbf{a}_2 + p_3 N_3 \mathbf{a}_3
\label{eq:T}
\end{equation}

\subsection{Hamiltonian in the Wannier basis} \label{subsec:hamiltonian}
From Eq.~(\ref{eq:Wannier_Rn}), one obtains the expression of the \textit{ab initio} Hamiltonian in the Wannier basis:
\begin{equation}
\mel{\mathbf{R}n}{\hat{H}}{\mathbf{R}^\prime n^\prime} = \frac{1}{N} \sum_{\mathbf{k}} e^{i\mathbf{k} \cdot (\mathbf{R}-\mathbf{R}^\prime)} \sum_{m=1}^{N_b} U_{mn}^{(\mathbf{k})*} \epsilon_{m\mathbf{k}} U_{mn^\prime}^{(\mathbf{k})}
\label{eq:Wannier_Hamiltonian}
\end{equation}
where $\epsilon_{m\mathbf{k}}$ are the Bloch states energies.
From these matrix elements, a Hamiltonian with finite range coupling is defined. It is meant to describe the infinite crystal instead of the ES, and it will be used for transport calculations and for interpolating the band structure at $\mathbf{k}$-points which are not in the MP grid. For this purpose, each Wannier function $\ket{\mathbf{R}n}$ is now interpreted as a localized function centered around $\mathbf{R}+\bar{\mathbf{r}}_n$, independent from its periodic replicas centered around $\mathbf{R}+\mathbf{T}+\bar{\mathbf{r}}_n$. Each $\ket{\mathbf{R}n}$ is only coupled to the $\ket{\mathbf{R}^\prime n^\prime}$ whose centers $\mathbf{R}^\prime+\bar{\mathbf{r}}_{n^\prime}$ are in the Wigner-Seitz cell centered around $\mathbf{R}+\bar{\mathbf{r}}_n$ and associated to the ES. In other words, $\mathbf{R}^\prime+\bar{\mathbf{r}}_{n^\prime}$ must be closer to $\mathbf{R}+\bar{\mathbf{r}}_n$ than any of its periodic replicas $\mathbf{R}^\prime+\bar{\mathbf{r}}_{n^\prime}+\mathbf{T}$.
This convention for defining the Hamiltonian is an available option in the \textsc{Wannier90} code. It implies, in the present study, that the maximum coupling range along the transport direction is $N_1 \times L/2$, where $L$ is the RS length.

The accuracy of this Hamiltonian improves rapidly when refining the MP grid of the \textit{ab initio} calculation. For the systems considered here, a $2 \times 4 \times 4$ grid is sufficient. The reason to use 2 $\mathbf{k}$-points along the transport direction is discussed in appendix \ref{appendix:accuracy}.

Obviously the Hamiltonian is modified when performing MB corrections. For $G_0 W_0$, the Bloch states are unchanged and only the eigenvalues $\epsilon_{m\mathbf{k}}$ are modified. Hence we keep the same Wannier functions and only recompute the Hamiltonian matrix elements of Eq.~(\ref{eq:Wannier_Hamiltonian}). For HSE and self-consistent COHSEX, since Bloch states are modified, Wannier functions must be re-optimized.

\subsection{Calculation of the tunneling conductance}
\label{subsec:conductance}

The Wannier Hamiltonian allows replacing the periodic repetition of the \textit{ab initio} supercell along the transport direction by semi-infinite Ag electrodes. It also allows interpolating the band structure on a finer grid of transverse $\mathbf{k}$-points. Indeed, the quantum transmission through the Ag/MgO/Ag junction strongly depends on the wavevector parallel to the interfaces, $\mathbf{k}^\parallel = k_2 \mathbf{g}_2 + k_3 \mathbf{g}_3$. The $4 \times 4$ grid used in our \textit{ab initio} simulations is not fine enough to resolve this dependence. Hence we consider a finer MP grid of $\mathbf{k}^\parallel$-points.
Wavefunctions are developped on the following basis, indexed by the lattice vector $\mathbf{R}^\perp = p_1 \mathbf{a}_1$ along the transport direction, the Wannier index $n$, and the transverse wavevector $\mathbf{k}^\parallel$:
\begin{equation}
\ket{\mathbf{R}^\perp n\mathbf{k}^\parallel} = \frac{1}{\sqrt{N^\parallel}} \sum_{\mathbf{R}^\parallel} e^{i\mathbf{k}^\parallel \cdot (\mathbf{R}^\parallel+\bar{\mathbf{r}}_n)} \ket{(\mathbf{R}^\perp+\mathbf{R}^\parallel) n}
\label{eq:nkp}
\end{equation}
where $N^\parallel$ is the number of $\mathbf{k}^\parallel$-points, and $\mathbf{R}^\parallel$ runs over transverse lattice vectors:
\begin{equation}
\mathbf{R}^\parallel = p_2 \mathbf{a}_2 + p_3 \mathbf{a}_3
\label{eq:Rt}
\end{equation}
These basis states are localized along the transport direction $\mathbf{a}_1$, and Bloch-like along the transverse directions. The Hamiltonian only couples states with the same transverse wavevector $\mathbf{k}^\parallel$ and its matrix elements are computed in terms of the matrix elements of Eq.~(\ref{eq:Wannier_Hamiltonian}):
\begin{equation}
H_{\mathbf{R}^\perp n \mathbf{R}^{\perp \prime} n^\prime}^{(\mathbf{k}^\parallel)} = \sum_{\mathbf{R}^{\parallel \prime}} e^{i\mathbf{k}^\parallel \cdot (\mathbf{R}^{\parallel \prime} +\bar{\mathbf{r}}_{n^\prime}-\bar{\mathbf{r}}_n)} \mel{\mathbf{R}^\perp n}{\hat{H}}{(\mathbf{R}^{\perp \prime}+\mathbf{R}^{\parallel \prime}) n^\prime}
\label{eq:Hkp}
\end{equation}

In some cases, the matrix elements will be neglected beyond a cut-off distance $L_{cut}$ along the transport direction. The cut-off condition reads:
\begin{equation}
\left| \left( \mathbf{R}^\perp +\bar{\mathbf{r}}_n - \mathbf{R}^{\perp \prime} -\bar{\mathbf{r}}_{n^\prime} \right) \cdot \hat{\mathbf{a}}_1 \right| > L_{cut}
\label{eq:Lcut}
\end{equation}
where $\hat{\mathbf{a}}_1$ is the unit vector along the transport direction.

We can now define the Hamiltonians of the central region and of the electrodes. The central region is defined as the RS and its $N_w \times N_w$ Hamiltonian reads:
\begin{equation}
H^C_{n n^\prime} = H_{\mathbf{0} n \mathbf{0} n^\prime}^{(\mathbf{k}^\parallel)}
\label{eq:HC}
\end{equation}
where we have ommited the transverse wavevector in $H^C$ to lighten the notations. The Hamiltonian of the periodic electrodes is extracted from Ag unit cells which are far enough from MgO so that its influence is screened.
We also need to define a cut-off distance beyond which the couplings between orbitals in Ag are negligible. A convergence study has shown that we can neglect the couplings between one cubic unit cell of Ag and its 4$^{th}$ neighbors and beyond. The remaining Hamiltonian blocks, schematized in Fig.~\ref{structure}, are extracted from the Hamiltonian (\ref{eq:Hkp}). More precisely, the diagonal block of the last Ag cubic cell, $H^{00}$, is made of the matrix elements:
\begin{equation}
H_{\mathbf{0} n \mathbf{0} n^\prime}^{(\mathbf{k}^\parallel)}
\label{eq:Hck}
\end{equation}
with $n$ and $n^\prime$ restricted to the orbitals of the last cubic cell. Similarly, the coupling blocks $H^{0i}$ between this cell and the first 3 cells of the right-side RS replica ($\mathbf{R}^\perp = \mathbf{a}_1$) are made of the matrix elements:
\begin{equation}
H_{\mathbf{0} n \mathbf{a}_1 n^\prime}^{(\mathbf{k}^\parallel)}
\end{equation}
with $n$ and $n^\prime$ restricted to the relevant cubic cells. The blocks $H^{0i}$ have dimension $28 \times 28$. They are used to build the Hamiltonians of the electrodes, which are block-tridiagonal, each block corresponding to 3 cubic cells. The right electrode Hamiltonian reads:
\begin{equation}
H^R =
\begin{pmatrix}
 D            & \tau   &        \\
 \tau^\dagger & D      & \ddots \\ 
              & \ddots & \ddots
\end{pmatrix}
\label{eq:HR}
\end{equation}
with the diagonal block:
\begin{equation}
D =
\begin{pmatrix}
H^{00}         &  H^{01}         &  H^{02} \\
H^{01 \dagger} &  H^{00}         &  H^{01} \\
H^{02 \dagger} &  H^{01 \dagger} &  H^{00}
\end{pmatrix}
\label{eq:D}
\end{equation}
and the coupling block:
\begin{equation}
\tau =
\begin{pmatrix}
H^{03} &  0      &  0      \\
H^{02} &  H^{03} &  0      \\
H^{01} &  H^{02} &  H^{03}
\end{pmatrix}
\label{eq:tau}
\end{equation}
The Hamiltonian of the left electrode is defined in a similar way, using the same blocks. $\tau$ also defines the coupling between the electrodes and the central region. The decimation algorithm of Lopez-Sancho \textit{et al.}\cite{Lopez-Sancho-1984,Lopez-Sancho-1985} is used to compute the electrodes self-energies. Full matrix inversion is used to compute the Green's function of the central region. Then the sum of transmission coefficients, $\mathcal{T}(\mathbf{k}^\parallel, E)$, is obtained from the trace formula\cite{Meir92}. It is averaged over transverse $\mathbf{k}^\parallel$-points to obtain the total sum of transmission coefficients per Ag cubic cell area:
\begin{equation}
\mathcal{T}(E) = \frac{1}{N^\parallel} \sum_{\mathbf{k}^\parallel} \mathcal{T}(\mathbf{k}^\parallel, E)
\label{eq:sumTk}
\end{equation}
In the following, a $10 \times 10$ MP grid of $\mathbf{k}^\parallel$-points is sufficient to reach convergence of $\mathcal{T}(E)$ around the Fermi energy.

Finally, we obtain the zero-bias conductance per Ag cubic cell area, at zero temperature:
\begin{equation}
\mathcal{G} = \mathcal{G}_0 \mathcal{T}(E_f)
\label{eq:conductance}
\end{equation}
where $\mathcal{G}_0 = 2e^2/h = 7.75 \times 10^{-5} \;  \Omega^{-1}$ is the quantum of conductance and $E_f$ is the Fermi energy, defined here as the charge neutrality point of the electrodes.

\subsection{Range of the Wannier Hamiltonian and cut-off distance}
\label{subsec:range}

\begin{figure}[h!]
\centering
\includegraphics[width=\linewidth]{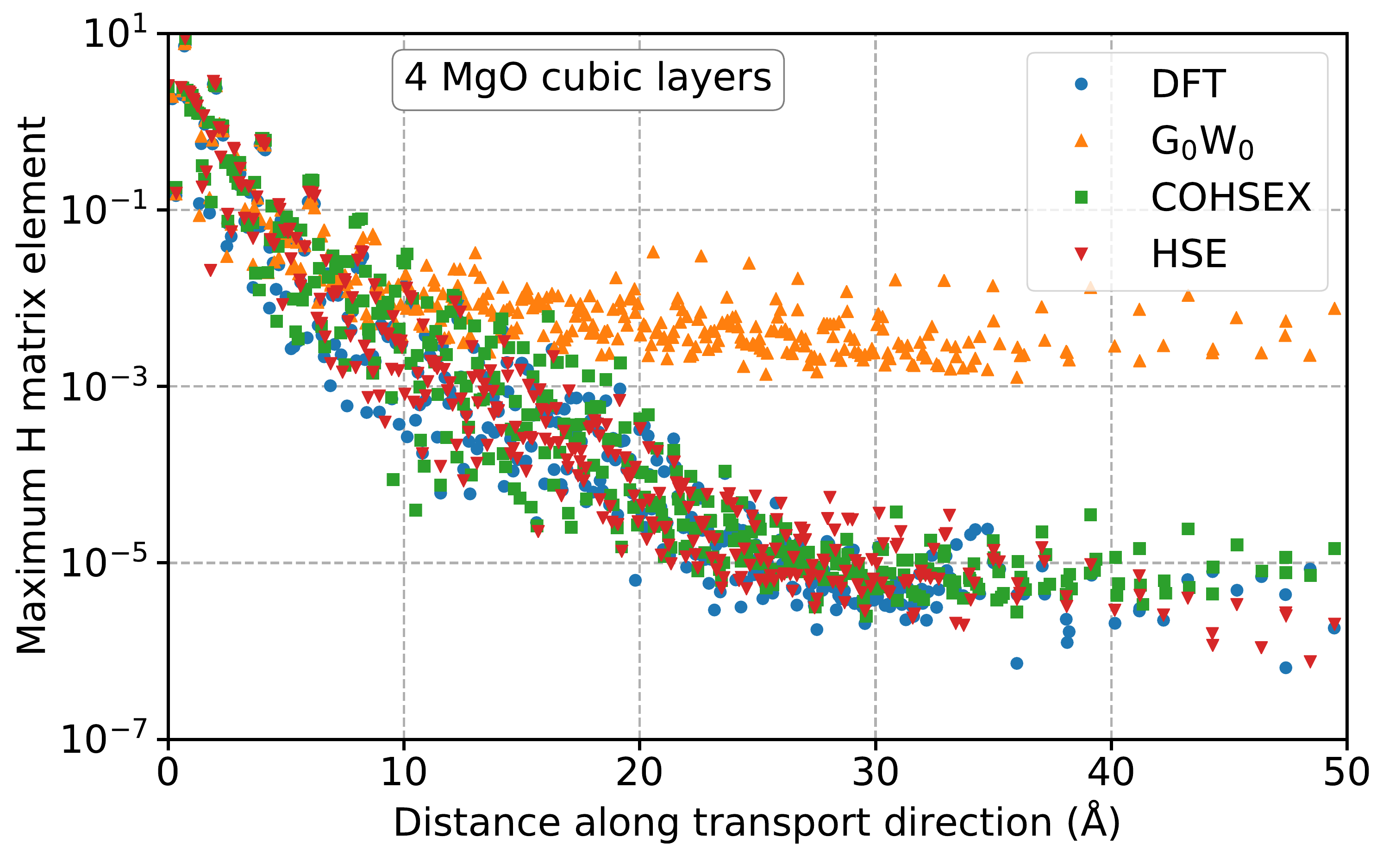}
\caption{Maximum modulus of the Wannier Hamiltonian matrix elements (\ref{eq:Hkp}) as a function of the distance between Wannier centers along the transport direction, as calculated by various approximations, here shown for the longest system (4Ag/4MgO/4Ag). The difference between DFT and $G_0W_0$ is only due to the change of eigenvalues in Eq.~(\ref{eq:Wannier_Hamiltonian}). On the contrary, for HSE and COHSEX, the matrix $U_{mn}^{(\mathbf{k})}$ also changes.}
\label{Hijgw}
\end{figure}

Fig.~\ref{Hijgw} shows how the maximum modulus of the Wannier Hamiltonian matrix elements (\ref{eq:Hkp}) decreases with the distance between Wannier centers along the transport direction, defined by the left-hand side of Eq.~(\ref{eq:Lcut}). The Hamiltonian is computed in DFT-PBE and with the three different MB corrections, using a $2 \times 4 \times 4$ MP grid.
The decay is exponential and similarly fast for all approximations up to a distance where it saturates to constant values. This distance is around 30 \angstrom \ to saturation values around $10^{-5}$ eV for all approximations except $G_0W_0$, which saturates before the others, around 13 \angstrom , to much higher values, around $10^{-2}$ eV. This is at first sight surprising since the Wannier functions are exactly the same than in DFT-PBE and well localized. However, the modification of the eigenvalues by the $G_0W_0$ correction are equivalent to a modification of the Hamiltonian which is local in energy, hence non-local in space. It induces long-range couplings between Wannier functions. This was already noticed by Ferretti \textit{et al.} \cite{Ferretti2012} who studied the impact of $G_0W_0$ corrections in conjugated polymers. These long-range couplings have a strong influence on the transmission spectra, which turn out to be noisy. An analysis detailed in appendix~\ref{appendix:accuracy} shows that, if the Hamiltonian accuracy is sufficient, transmission spectra should be nearly insensitive to long-range couplings, which could be safely discarded. We conclude that the high $G_0W_0$ long-range couplings may be not physical, and we decide to discard them beyond a cut-off distance $L_{cut} = 13$ \angstrom , after which the $G_0W_0$ matrix elements become higher than in the other approximations.
However, in the following, one should keep in mind that $G_0W_0$ spectra are subject to caution, and that $G_0W_0$ does not seem to be a reliable method to compute the transport properties of tunnel junctions.
\section{Results}
\label{sec:results}

\subsection{Transmission spectra in the DFT approximation}
\begin{figure}[h!]
\centering
\includegraphics[width=\linewidth]{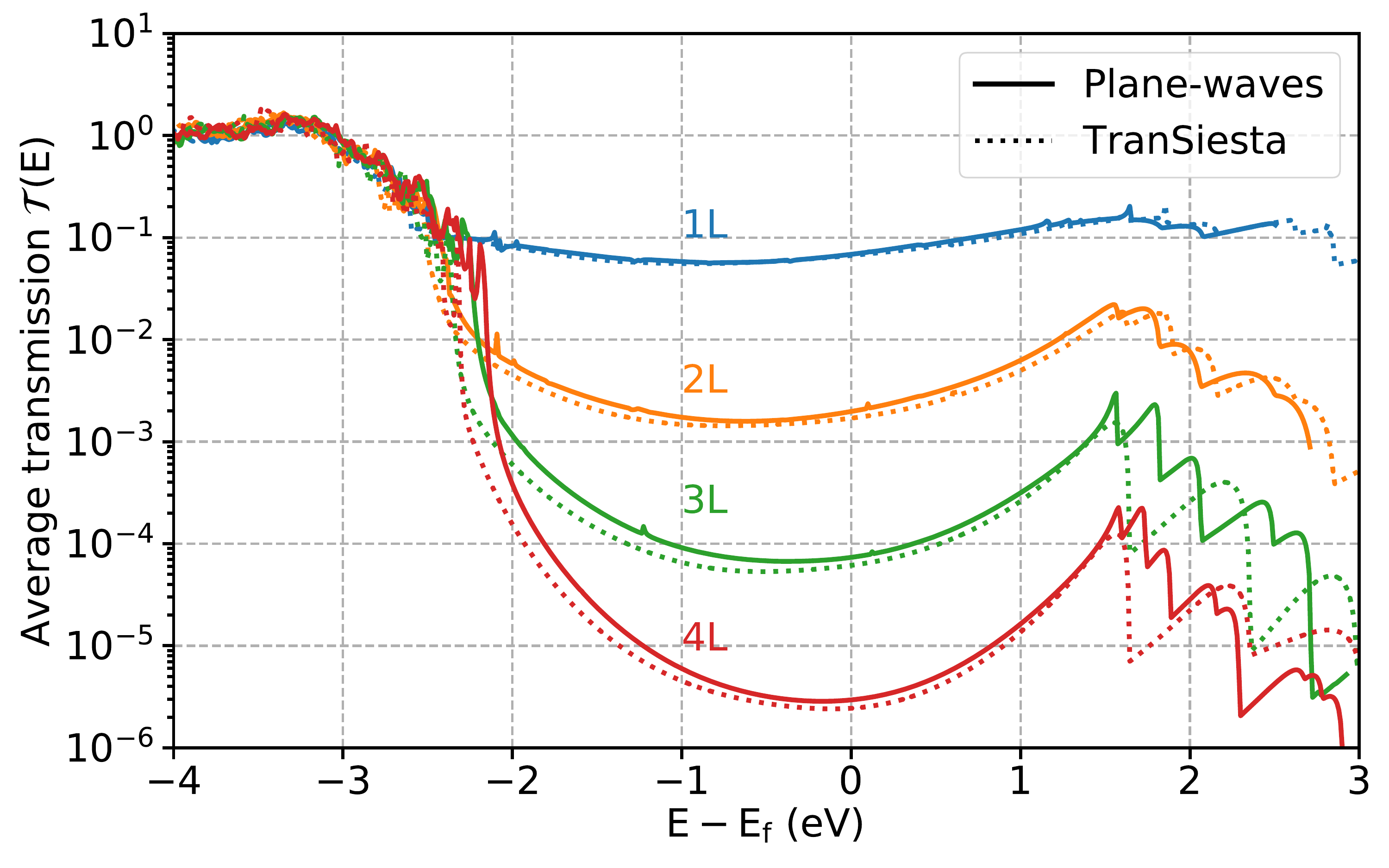}
\caption{Transmission spectra $\mathcal{T}(E)$ through MgO barriers of different thicknesses, from 1 cubic layer (1L) to 4 cubic layers (4L). The calculations are done in the DFT-PBE approximation. The results obtained by the plane-waves/Wannier method are compared to those obtained by \textsc{TranSiesta}.}
\label{DFT-TS_table}
\end{figure}

The transmission values computed at the level of DFT-PBE are expected to be consistent with existing codes based on localized functions basis set. Such comparison is interesting to see if the results are robust with respect to the basis choice.
Fig.~\ref{DFT-TS_table} shows the transmission spectra $\mathcal{T}(E)$ calculated for different MgO thicknesses, in the DFT-PBE approximation. As expected, the plane-waves/Wannier results are in very good agreement with those obtained with \textsc{TranSiesta}.

We will now provide an interpretation of the transmission curves.
At low energies we are in the valence band (VB) of MgO, 
and the transmission is high and independent of the MgO thickness. The VB edge lies at about -2.2 eV below $E_f$. Beyond the VB, in the MgO band-gap, we enter into the tunneling regime and the transmission decreases exponentially with the thickness.
The values of the zero-bias conductance are given in Table~\ref{conductance_table} and its exponential decay with the MgO thickness is shown in Fig.~\ref{conductance_both}.

The analysis, not shown here, of the $\mathbf{k}^\parallel$-resolved transmission $\mathcal{T}(\mathbf{k}^\parallel, E_f)$ shows that the main contribution comes from transverse wavevectors close to zero. This is similar to what is obtained in DFT+Landauer simulations of Fe/MgO/Fe magnetic tunnel junctions with parallel magnetizations \cite{Butler_abinitio_Fe-MgO_2001}. The decay rate of the conductance with the MgO thickness is also in very good agreement with Ref.~[\onlinecite{Butler_abinitio_Fe-MgO_2001}].
The decay is associated to evanescent states in the gap of MgO which have $\Delta_1$ symmetry at $\mathbf{k}^\parallel = \mathbf{0}$ \cite{Butler_abinitio_Fe-MgO_2001}. The first conduction band of MgO has the same symmetry. However, the conduction band edge, which is expected to be at about 2.5 eV above $E_f$, is not visible on the transmission plot. This is due to the transition, at $E > E_f+1.6$ eV, to an Ag band with a different symmetry, which cannot be transmitted through the $\Delta_1$ states of MgO.

\subsection{Effect of MB corrections on the tunneling transmission}
\begin{figure*}[t]
\centering
\includegraphics[width=0.5\linewidth]{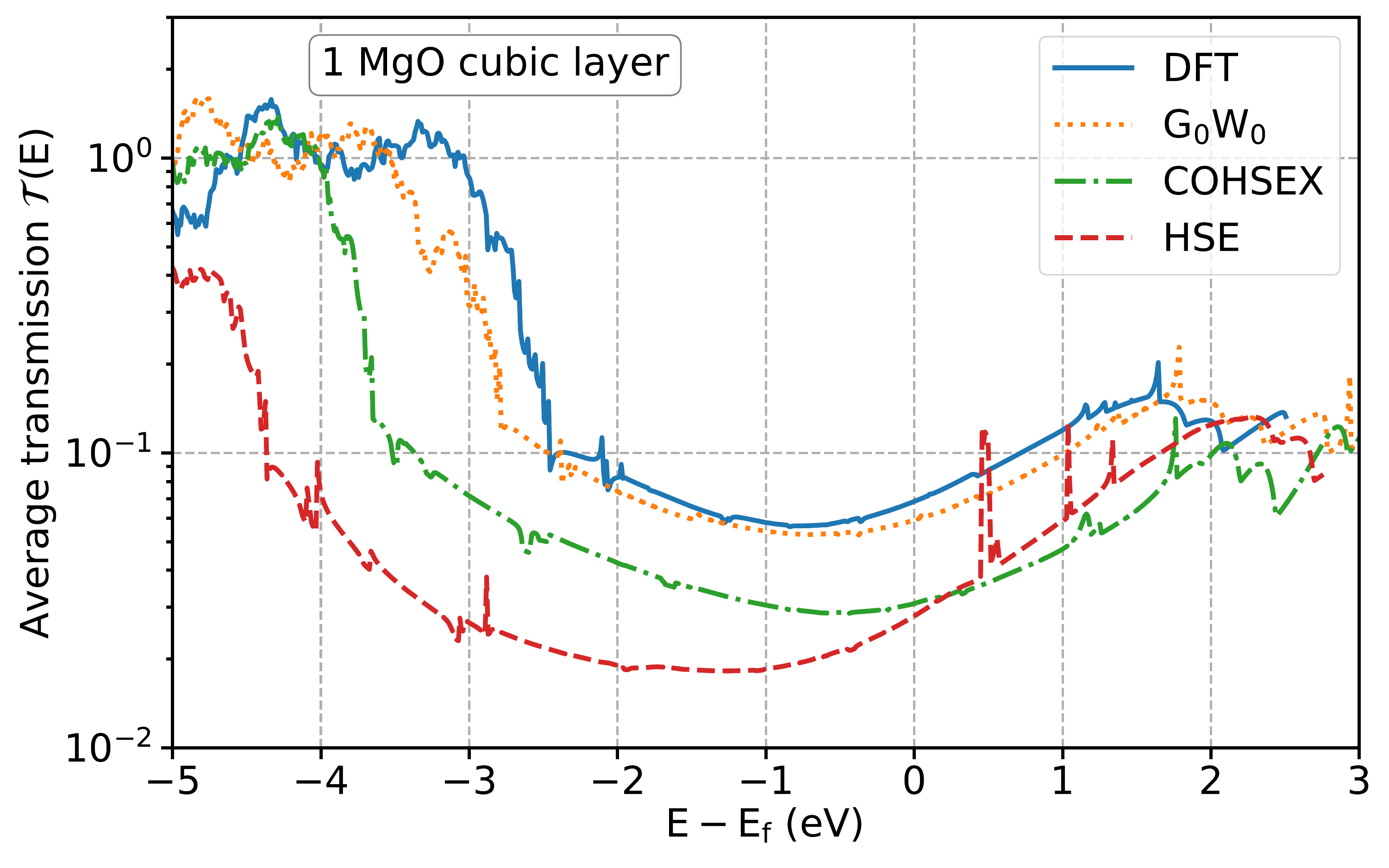}\hfill
\includegraphics[width=0.5\linewidth]{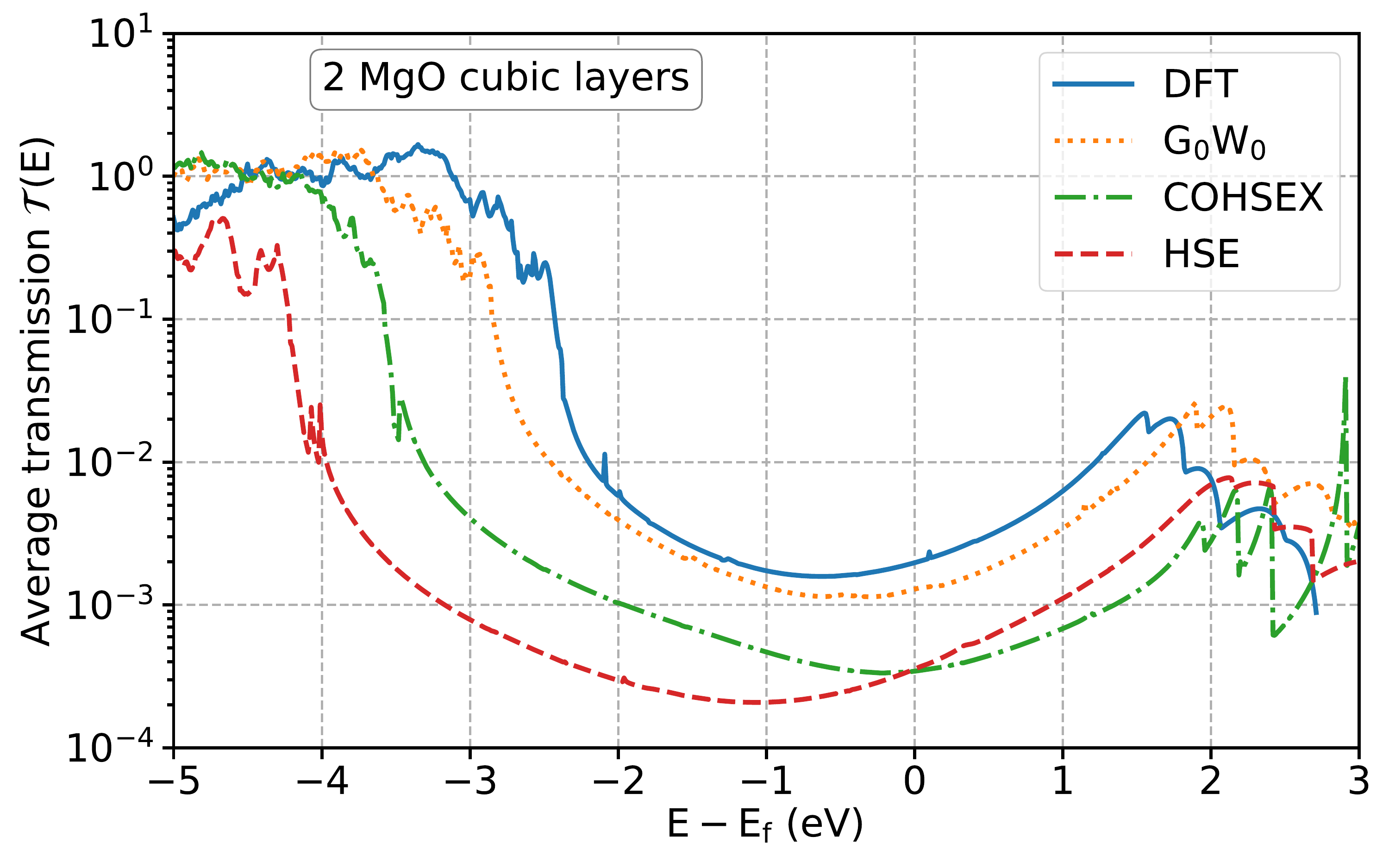}\par\medskip
\includegraphics[width=0.5\linewidth]{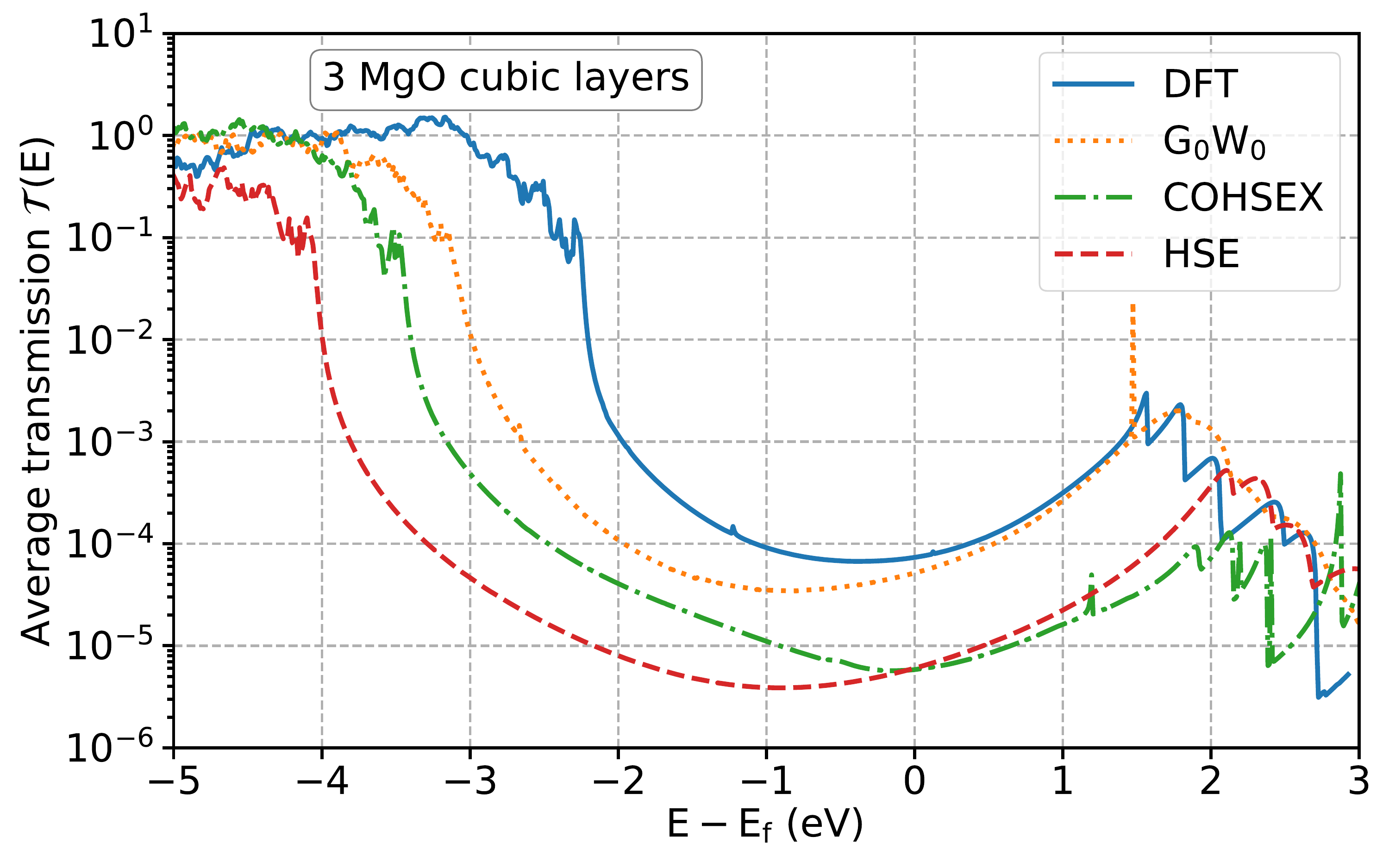}\hfill
\includegraphics[width=0.5\linewidth]{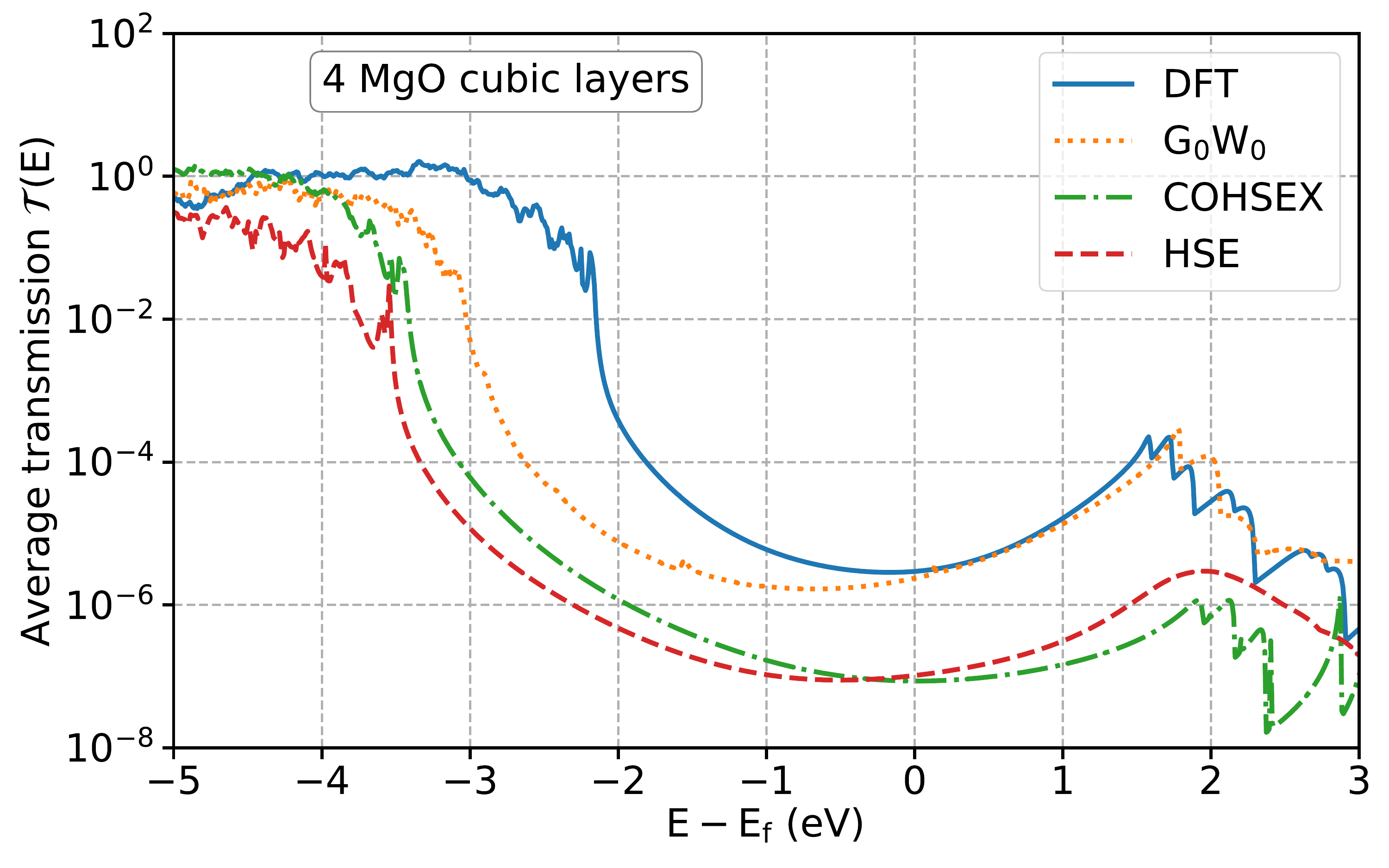}
\caption{Transmission spectra calculated by various approximations, for different MgO thicknesses (1 to 4 cubic layers). All the $G_0W_0$ transmission functions, and the HSE and COHSEX ones for the thickest structure (4MgO), are calculated by applying a Hamiltonian cut-off distance $L_{cut}$ of 13 \angstrom \  (see section~\ref{subsec:range} and appendix~\ref{appendix:accuracy}).}
\label{transmissions}
\end{figure*}

Figure \ref{transmissions} shows how $G_0W_0$, self consistent COHSEX (simply referred as COHSEX in this section) and HSE modify the DFT transmission function, for the 4 structures with increasing MgO thickness, from 1 to 4 cubic layers. For the lowest transmission spectra, that is HSE and COHSEX applied to the thickest MgO structure, the long range Hamiltonian couplings were discarded by applying a cut-off distance $L_{cut} = 13$ \angstrom . This has eliminated some moderate numerical noise occurring at these very low tunnel transmissions below $10^{-7}$. This point is detailed in appendix \ref{appendix:accuracy}.

We expect to observe a reduction of the transmission spectra with respect to the DFT ones.
This is confirmed for COHSEX and HSE, which for the 3 and 4 MgO structures reduce the $\mathcal{T}(E_f)$ value of more than 1 order of magnitude. 
For a more complete analysis it is interesting to compare our results to the experimental results of Gangineni \textit{et al.} [\onlinecite{Gangineni_Fe-MgO_2014}]. In this work, measurements of the tunneling conductance have been performed on Fe/MgO/Fe junctions at different MgO thicknesses, up to 4 cubic cells. The experimental conductances for parallel magnetizations of the Fe electrodes have been then compared to the theoretical values obtained by Butler \textit{et al.} [\onlinecite{Butler_abinitio_Fe-MgO_2001}] with the DFT+Landauer approach. For 4 MgO cubic cells, the deviation between experiment and theory turns out to be between 1 and 2 orders of magnitude (Fig.~3 of Ref.~[\onlinecite{Gangineni_Fe-MgO_2014}]). This is comparable to the deviation observed in our Ag/4MgO/Ag system between DFT and COHSEX or HSE. It indicates that the correction introduced by COHSEX and HSE goes in the right direction to improve the agreement with experiments. This comparison is qualitative since the electrodes considered in the experiment are made of Fe. But it remains physically relevant since, for both Ag electrodes and Fe electrodes with parallel magnetizations, the exponential decay of the conductance with the MgO thickness is related to evanescent states of symmetry $\Delta_1$ in MgO. 

\begin{table}[h!]
\caption{Zero-bias tunneling conductance (in $\Omega^{-1}.\mu \mathrm{m}^{-2}$) through a MgO barrier of different thickness, from 1 to 4 cubic cells, with Ag electrodes, as calculated by various approximations. The conductances are evaluated at zero temperature by Eq.~(\ref{eq:conductance}). Their values at room temperature, not given here, do not differ by more than a few percents.}
\begin{ruledtabular}
\begin{tabular}{lcccc}
 Method     & 1 MgO & 2 MgO & 3 MgO & 4 MgO \\
 \hline
 DFT-PBE         & $31$  &  $8.9 \times 10^{-1}$ & $3.3 \times 10^{-2}$ & $1.5 \times 10^{-3}$\\
 PBE + $G_0W_0$  & $27$  &  $5.8 \times 10^{-1}$ & $2.3 \times 10^{-2}$ & $1.1 \times 10^{-3}$\\
 COHSEX          & $14$  &  $1.6 \times 10^{-1}$ & $2.6 \times 10^{-3}$ & $3.9 \times 10^{-5}$\\
 HSE             & $13$  &  $1.6 \times 10^{-1}$ & $2.7 \times 10^{-3}$ & $4.7 \times 10^{-5}$\\
\end{tabular}
\end{ruledtabular}
\label{conductance_table}
\end{table}

\begin{figure}[h!]
\centering
\includegraphics[width=\linewidth]{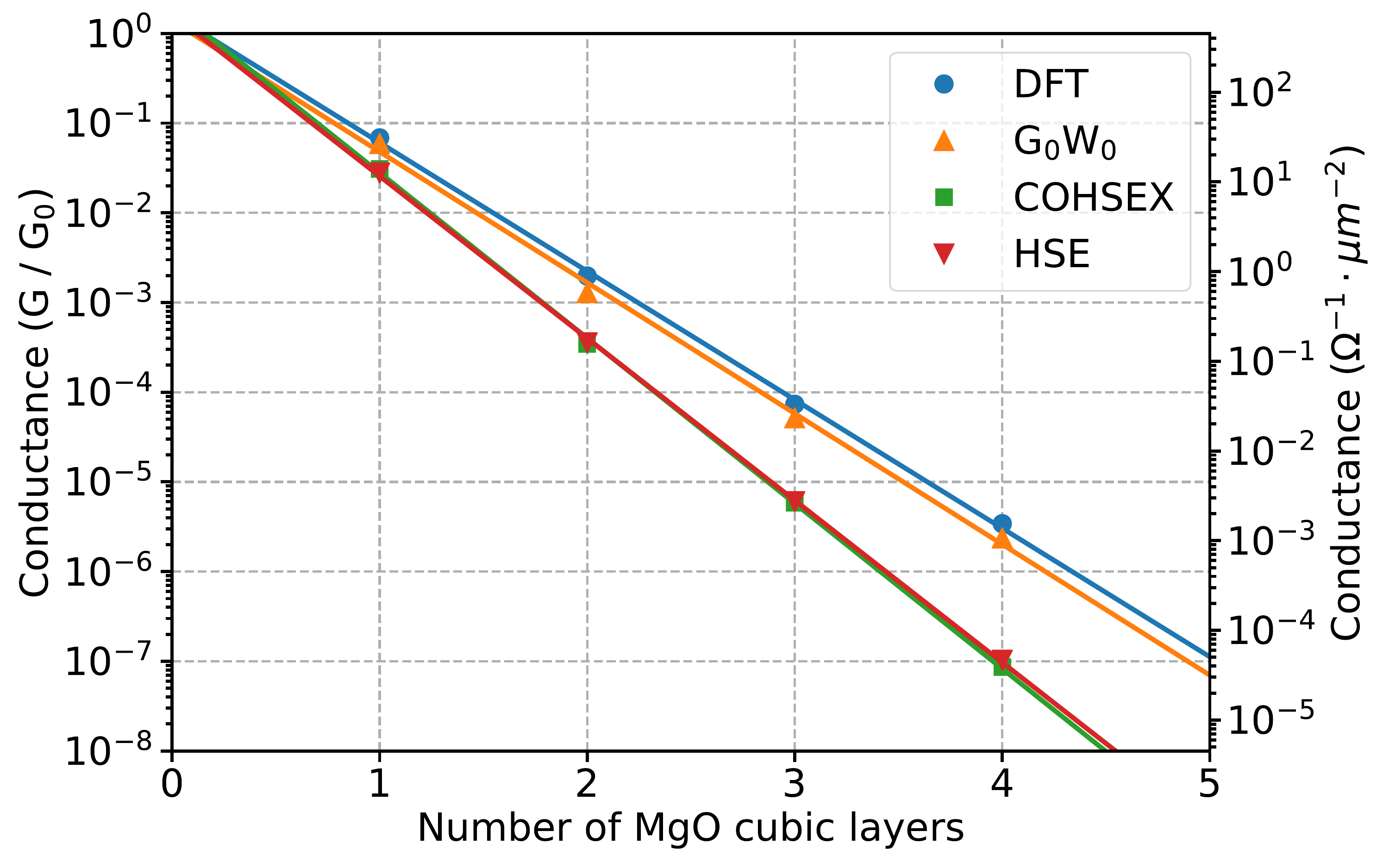}
\caption{Values of table \ref{conductance_table} in logarithmic scale with respect to the number of MgO layers in the central region. The left y-axis shows values in unit of $G_0 = 7.75 \times 10^{-5} \; \Omega^{-1}$, that is $\mathcal{T}(E_f)$, while the right y-axis shows values in $\Omega^{-1}.\mu m^{-2}$. The straight lines are linear regressions on the data logarithm. The decreasing of the conductance is thus exponential.}
\label{conductance_both}
\end{figure}

The values of zero-bias conductance (at zero temperature), that is $\mathcal{T}(E_f)$, are shown in table \ref{conductance_table} for the four structures and four approximations. Fig.~\ref{conductance_both} shows that the decay with the MgO thickness is exponential, with a net separation between DFT/$G_0W_0$ and HSE/COHSEX. 
For DFT the decay rate is very close to what was found in Ref.~[\onlinecite{Butler_abinitio_Fe-MgO_2001}]. We have $\mathcal{G} \propto \exp(-\kappa \times n_{MgO})$, where $n_{MgO}$ is the number of MgO cubic cells and $\kappa \simeq 3.3$. $\kappa$ increases for HSE and COHSEX to values around 4.2. This entails that the overestimation error of the DFT+Landauer approach becomes more important in higher tunneling regimes. 
The decay rate obtained in HSE and COHSEX ($\kappa \simeq 4.2$) is also in good agreement with the rate obtained for the A sample in Ref.~[\onlinecite{Gangineni_Fe-MgO_2014}]. On the contrary the decay rate obtained in DFT ($\kappa \simeq 3.3$) is closer to the B sample (Fig.~3 of Ref.~[\onlinecite{Gangineni_Fe-MgO_2014}]).

The fact that $G_0W_0$ does not provide a significant reduction of transmission and conductance (Figs~\ref{transmissions} and \ref{conductance_both}) leads to another interesting analysis.
The DFT systematical underestimation of the insulators band-gap\cite{Perdew_bandgap} is suspected to be the main cause of the overestimation of the tunneling transmission. This is inferred by assuming that the band offsets between the Fermi level $E_f$ of the electrodes and the valence and conduction band edges of MgO are both underestimated by DFT. This underestimation of the energy barriers should lead to an overestimation of tunneling. MB corrections on eigenvalues are thus expected to give more accurate conductance values. 
\begin{table}[h]
\caption{
Electronic band-gap of bulk MgO, calculated by various approximations. The lattice parameter equals 4.212 \angstrom, which is the experimental value at room temperature\cite{Wyckoff}. Values are in eV.}
\begin{ruledtabular}
\begin{tabular}{lcccc}
 Method & This work & Other works \\
 \hline
 DFT-PBE & 4.72 & 4.68\cite{Pasquarello-MgO-2015}; 4.45\cite{Pasquarello-MgO-2012} \\
 PBE + $G_0W_0$ & 7.44 & 7.08\cite{Pasquarello-MgO-2015}; 7.41\cite{Erratum-Pasquarello-MgO-2013} \\
 COHSEX & 9.37 & - \\
 HSE & 7.77 & 7.67\cite{HSEstudyMgO} \\
 Experimental & - &  7.67 - 7.83\cite{HSEstudyMgO} \\
\end{tabular}
\end{ruledtabular}
\label{MgO-gap}
\end{table}
Table~\ref{MgO-gap} shows the electronic band-gap of bulk MgO calculated in different approximations: DFT-PBE, $G_0 W_0$, COHSEX, and HSE.
$G_0 W_0$ gives a band-gap close to the experimental values, whereas COHSEX, as expected, overestimates the band-gap.
Our observation is that the decrease of the tunneling transmission is not simply correlated to the opening of the MgO band-gap. In the $G_0 W_0$ approximation, the zero-bias conductances remain close to the DFT values, despite the band-gap opening from 4.72 to 7.44 eV. The HSE approximation gives similar gap opening but a smaller transmission.
The COHSEX approximation gives conductances very close to the HSE ones, while the band-gap is much larger. The band offset between $E_f$ and the VB edge of MgO is also not simply correlated to the band-gap opening. In particular, the offset is larger in HSE than in COHSEX. The HSE offset is around 4 eV, which is close to the experimental value obtained by scanning tunneling spectroscopy on thin MgO layers deposited on an Ag substrate \cite{Schintke2001}.

These results point to the fact that it is not the correction of the band-gap and energies which drives the reduction of the zero-bias conductance, but rather the correction of wavefunctions. Indeed, $G_0W_0$, which keeps the Kohn-Sham eigenfunctions unmodified, does not provide any significant correction to the DFT zero-bias conductance, while HSE and COHSEX, which update also eigenfunctions, provide significant and similar corrections to the zero-bias conductance, despite the different MgO band-gap.
\begin{figure}[h!]
\centering
\includegraphics[width=\linewidth]{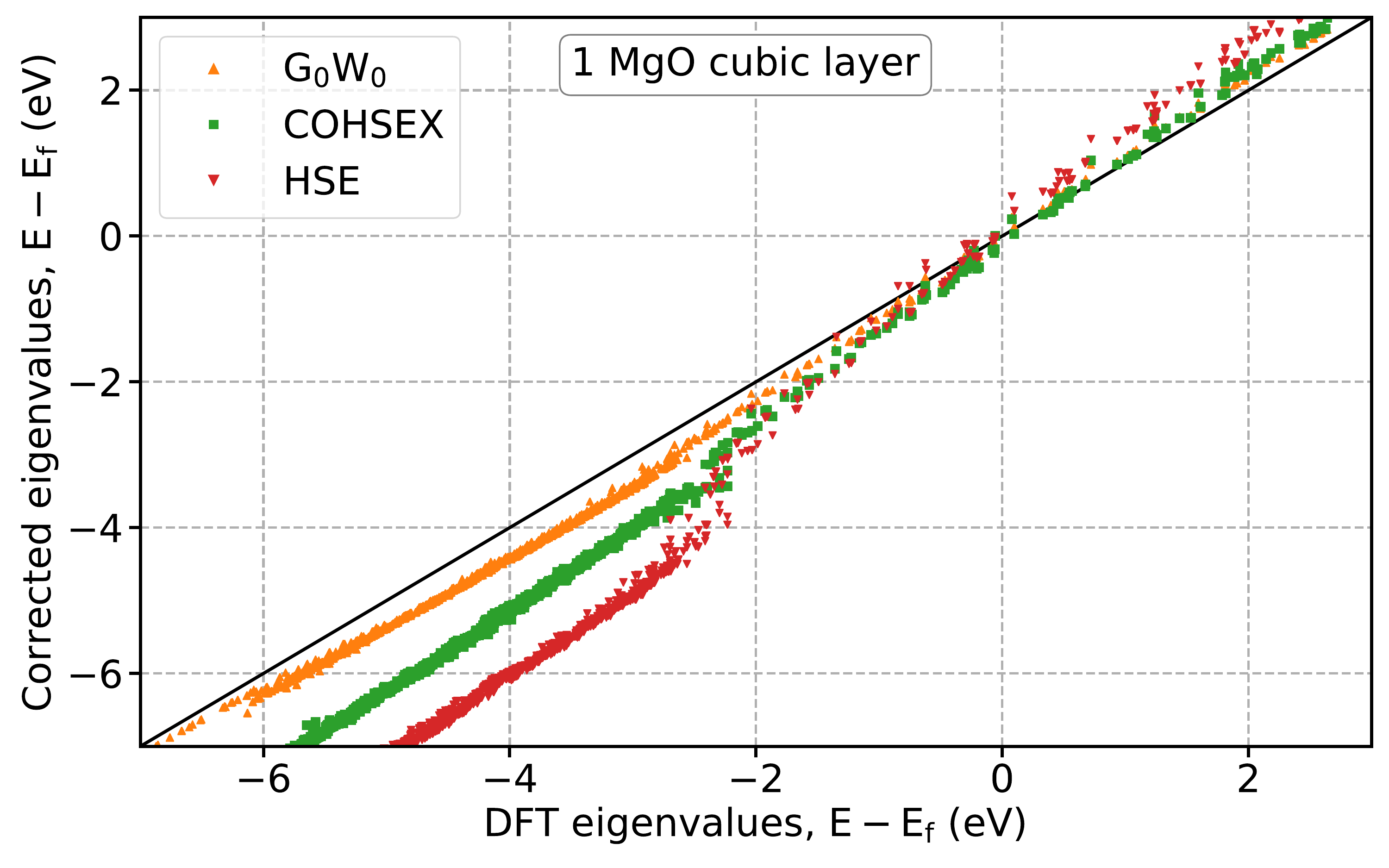}
\caption{Quasiparticle energies as function of the DFT-PBE eigenvalues shown for the Ag/1MgO/Ag structure.} 
\label{eigenvalues}
\end{figure}
Fig.~\ref{eigenvalues}, which shows how DFT eigenvalues of the 4Ag/1MgO/4Ag supercell are modified by the different MB corrections, also supports this argument. Eigenvalues close to the Fermi energy correspond to wavefunctions mainly located in Ag, with evanescent tails in MgO. Eigenvalues of a metal, which form a continuous spectrum, are not very sensitive to QP corrections, and remain almost unmodified with respect to DFT.
Modifications occur only far away from the Fermi energy and for Bloch states with contributions which start to come from Mg and O electrons.
This means that a change in the transmission function around the Fermi energy is necessarily caused by the modification of eigenfunctions. This follows by the fact that the trace formula used to compute the transmission is local in energy. In other words, the transmission at energy $E$ can be obtained by building diffusion eigenstates made of incident, reflected, and transmitted waves. These states are linear combinations of the \textit{ab initio} Bloch eigenfunctions at energy $E$. Hence a modification of these wavefunctions is needed to modify the DFT transmission close to $E_f$, since eigenvalues remain basically unmodified.
This point seems to have some analogies with the conclusions of Ferretti~\textit{et al.}\cite{Ferretti2012}, who computed the decay of evanescent states in conjugated polymers under various DFT+MB approximations.

A final comment is worth on the comparison between COHSEX and HSE.
The agreement of HSE with COHSEX on the 0-bias conductance shown in Fig.~\ref{conductance_both} and Table~\ref{conductance_table} might be fortuitous. Indeed, COHSEX and HSE differ on all other quantities, e.g. eigenvalues, band-gaps, so that it looks unlikely that HSE and COHSEX could provide the same eigenfunctions, even though if only those located at the Fermi energy.

\section{Conclusion}

The tunneling transmission spectra and zero-bias conductances of Ag/MgO/Ag junctions have been computed using DFT-based \textit{ab initio} calculations. A methodology combining plane-waves DFT and transformation to a Wannier basis set has been used to compute the quantum transmission. Since DFT systematically underestimates the band-gap of insulators, it is expected that the tunneling transmission in the coherent regime is overestimated. This should be improved by MB corrections. For this purpose, the impact of different MB corrections to DFT has been evaluated. Our conclusion is that MB corrections of energies and band offsets have nearly no impact on the zero-bias conductance if the DFT eigenfunctions are not updated. The conductance is significantly reduced only if the eigenfunctions are updated in the MB correction scheme. Ferretti \textit{et al.} \cite{Ferretti2012} gave a similar conclusion about the decay lengths of evanescent states in the band-gap of conjugated polymers. Here we have reinforced this conclusion by evaluating MB corrections on a bulk tunneling heterojunction, which was to our knowledge not done in the literature. Indirect comparison of our simulations with experiments (Gangineni \textit{et al.}\cite{Gangineni_Fe-MgO_2014}) indicate that MB corrections goes toward more accurate tunneling conductances.

HSE and self-consistent COHSEX seem the most practical options for performing MB corrections of the tunneling transmission, since these corrections are available in many DFT codes. Self-consistent $GW$ would be in principle more accurate but it is more computationally demanding. An interesting but ambitious perspective would be to include MB corrections directly inside the Non Equilibrium Green's Functions (NEGF) formalism, starting from a DFT Hamiltonian in a localized basis set (Wannier functions or atomic orbitals). This should correct the band-gap and zero-bias conductance, and would allow considering finite bias voltages and electron-phonon interaction.

\appendix

\section{Accuracy of the Wannier Hamiltonian and cut-off distance}
\label{appendix:accuracy}

\begin{figure}[h!]
\centering
\includegraphics[width=\linewidth]{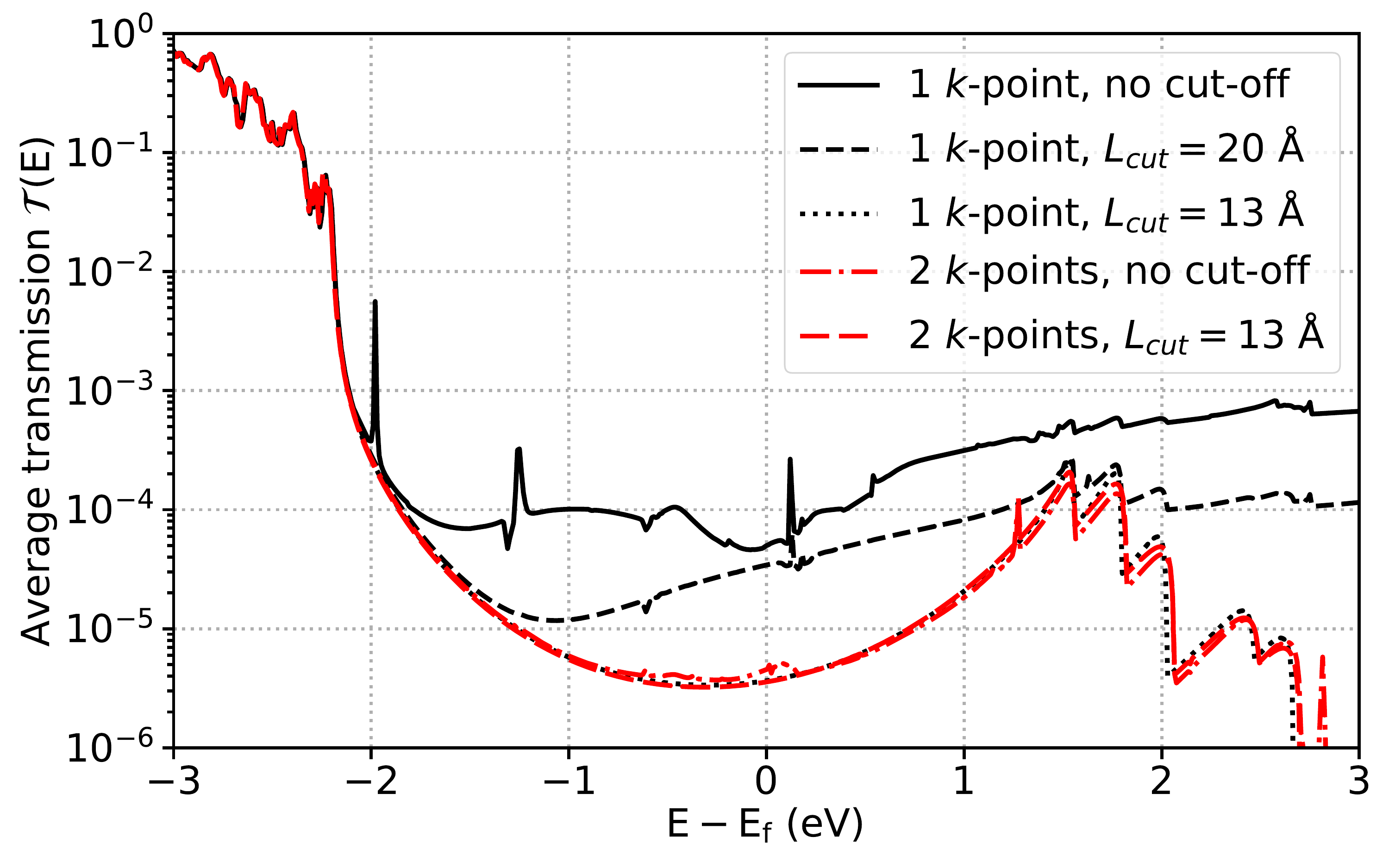}
\caption{Transmission spectrum of the 4Ag/4MgO/4Ag structure in DFT-PBE, using the plane-waves/Wannier method. If a single $\mathbf{k}$-point is used along the transport direction, the spectrum depends on the cut-off distance $L_{cut}$. If 2 $\mathbf{k}$-points are used, the spectrum becomes nearly independent on $L_{cut}$ down to $\sim 13$~\angstrom . Moreover, the spectra with 1 or 2 $\mathbf{k}$-points are nearly identical at $L_{cut} = 13$~\angstrom .}
\label{fig:transmission_lcut}
\end{figure}

During the first tests of the plane-waves/Wannier methodology, we used a $1 \times 4 \times 4$ MP grid in the \textit{ab initio} calculations. Using a single $\mathbf{k}$-point $\Gamma$ along the transport direction seemed sufficient, given the long length of the considered supercells. However, for the thickest oxide, we have encountered accuracy issues in transport calculations. Fig.~\ref{fig:transmission_lcut} shows the DFT transmission spectra obtained with different values of the cut-off distance $L_{cut}$. The couplings at distances around $20$ \angstrom \ strongly modify the tunneling transmission, and there is no evident justification for discarding them. The problem was solved by using a $2 \times 4 \times 4$ MP grid. Fig.~\ref{fig:transmission_lcut} shows that the DFT transmission spectrum is then nearly insensitive to the cut-off. This proves that the couplings at distances beyond $13$ \angstrom \ have negligible impact on the transmission and can be safely discarded.

\begin{figure}[h!]
\centering
\includegraphics[width=\linewidth]{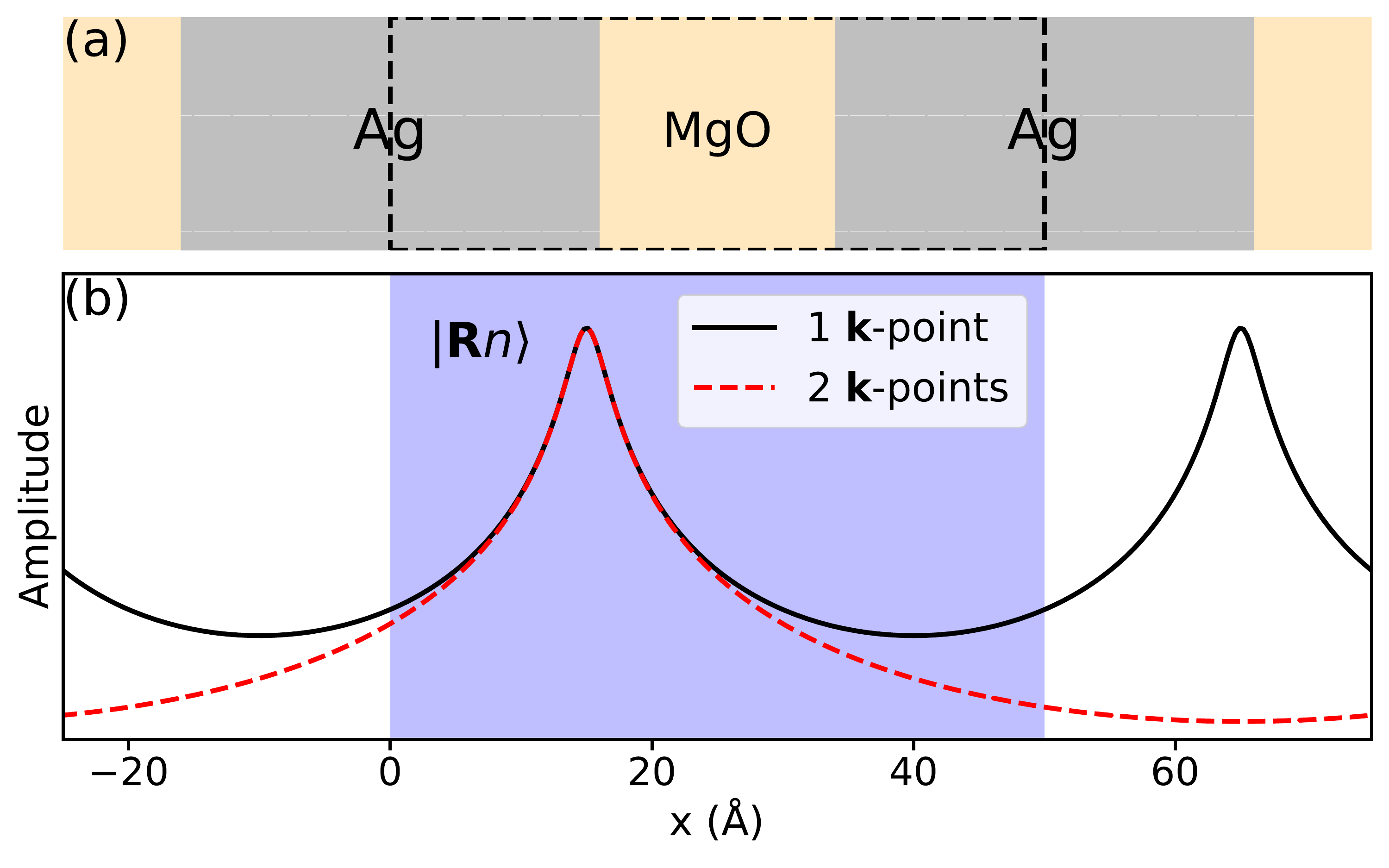}
\caption{(a) 4Ag/4MgO/4Ag supercell, delimited by a dashed line, and portions of its periodic replicas. (b) Sketch of a Wannier functions computed with 1 and 2 $\mathbf{k}$-points along the transport direction (log scale). By using one single $\mathbf{k}$-point the period of the Wannier function equals the supercell length $L$. The Hamiltonian matrix element between the two functions is impacted by the tails of the periodic replicas, leading to an overestimate of the direct coupling between Ag electrodes. By using 2 $\mathbf{k}$-points, the Wannier replicas are moved away at distance $2L$ (not shown here) and have much less influence.}
\label{fig:replicas}
\end{figure}

\begin{figure}[h!]
\centering
\includegraphics[width=\linewidth]{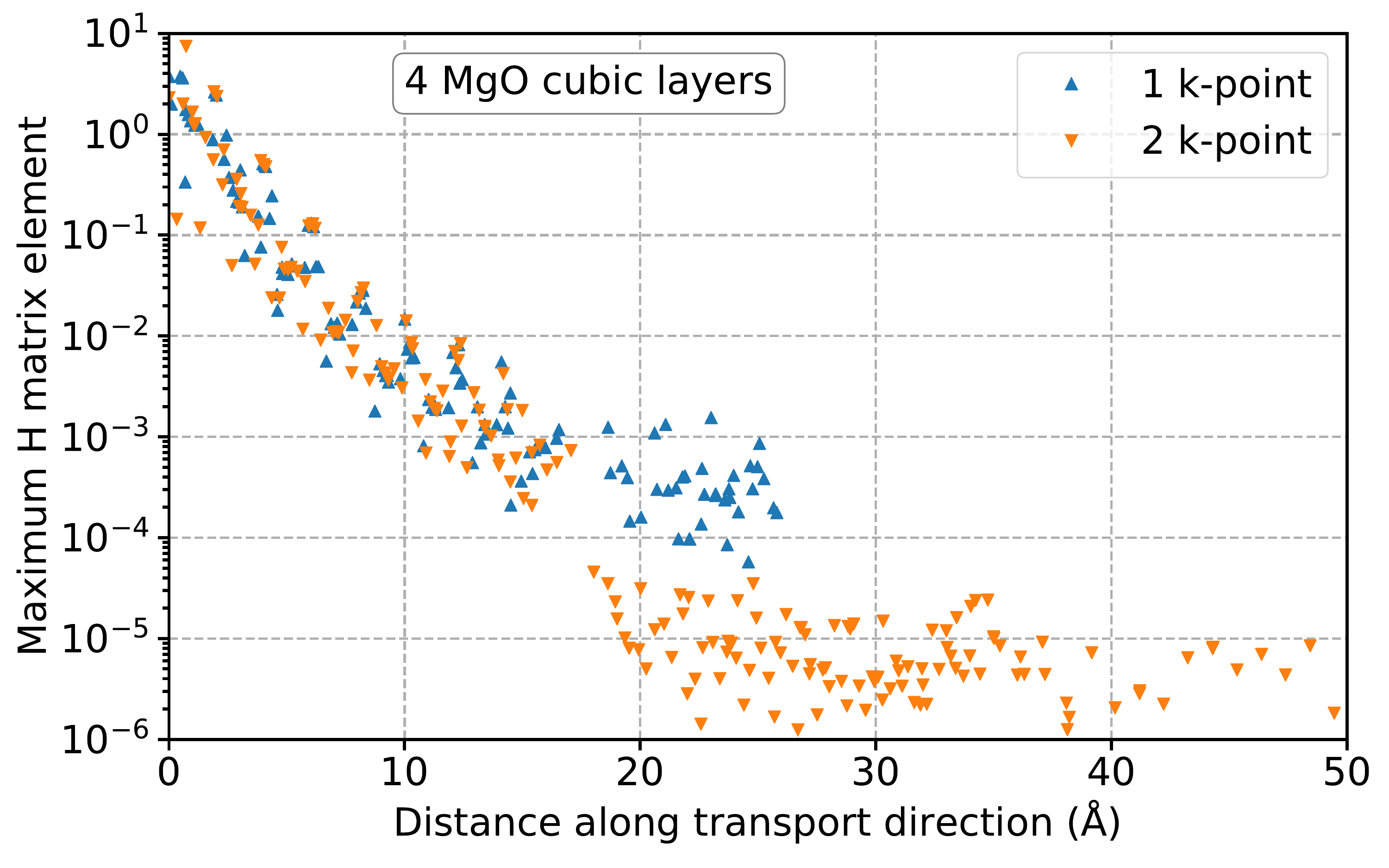}
\caption{Decay of the Hamiltonian couplings between Ag Wannier functions with the distance. The calculation is done on a 4Ag/4MgO/4Ag supercell in the DFT-PBE approximation, using either 1 or 2 $\mathbf{k}$-points along the transport direction.}
\label{fig:Hij_1vs2kp}
\end{figure}

The explanation is that the distant couplings are impacted by the periodicity of the Wannier functions along the transport direction. When using a single $\mathbf{k}$-point, this period equals the length $L$ of the supercell, and the influence of the replicas can be strong, as illustrated in Fig.~\ref{fig:replicas}. When using 2 $\mathbf{k}$-points, the period of the Wannier functions is $2L$ and the periodic replicas have much less impact. 
The coupling of the Wannier function $\ket{\mathbf{R}n}$ in Fig.~\ref{fig:replicas} (b) with another function $\ket{\mathbf{R}^\prime n^\prime}$ is defined as an integral over the ES. Hence, in the 1 $\mathbf{k}$-point case, the coupling calculation requires to integrate the black plain line over the blue area in Fig.~\ref{fig:replicas} (b), whereas it requires to integrate the red dashed line over the whole $x$-axis range when using 2 $\mathbf{k}$-points. It is evident that, in the first case, the impact of the closer periodic replicas leads to an overestimation of the coupling. This overestimation can be negligible for strong, short-distance couplings, but dominant for weak, long-distance ones.

This interpretation is supported by Fig.~\ref{fig:Hij_1vs2kp}, which shows the decrease of Ag--Ag DFT Hamiltonian couplings with the distance between Wannier centers. This is the same analysis as in Fig.~\ref{Hijgw}, but restricted to Wannier functions belonging to the Ag electrodes. The couplings beyond $17$ \angstrom \ are direct tunneling Ag--Ag couplings through MgO. These couplings, which are expected to be very weak, are much larger with 1 $\mathbf{k}$-point than with 2 $\mathbf{k}$-points. This overestimation when using 1 $\mathbf{k}$-point is due to the effect of replicas, which dominates the true, physical, weak couplings. On the contrary, for short-distance strong Ag--Ag couplings inside the electrode ($< 17$ \angstrom), the effect of replicas is negligible. The overestimated direct Ag--Ag couplings beyond $17$ \angstrom \ are non-physical, and introduce some noise in the transmission spectrum.

Hence, to ensure a good accuracy, we have performed all \textit{ab initio} calculations of this study with a $2 \times 4 \times 4$ MP grid. In the DFT-PBE, HSE, and self-consistent COHSEX approximations, the transmission is nearly insensitive to an increase of $L_{cut}$ beyond $13$ \angstrom. Only for the thickest oxide, where Ag--Ag direct couplings are extremely weak, we observe small fluctuations in the tunneling transmission when $L_{cut}$ approaches the maximum coupling distance $L \simeq 50$ \angstrom. Such fluctuations are visible in Fig.~\ref{fig:transmission_lcut}, when using 2 $\mathbf{k}$-points and no cut-off. These very distant couplings are related to the periodicity $2L$ of the Wannier functions and must be discarded. Hence we have used a cut-off distance of $13$ \angstrom \ to suppress these fluctuations when necessary. This cut-off distance is shorter than the MgO thickness, thus all the Ag--Ag couplings through MgO are forced to zero.

Using 2 $\mathbf{k}$-points does not solve the problem of long range interactions in the $G_0W_0$ approximation. As discussed in section~\ref{subsec:range}, these interactions are intrinsic to the method.

\end{document}